\documentclass[prl,amsmath,amssymb,twocolumn,showpacs,floatfix]{revtex4}
\usepackage{graphicx}
\usepackage{bm}

\begin{document}

\title{Inhomogeneous Magnetic-Field Response of YBa$_2$Cu$_3$O$_y$ and La$_{2-x}$Sr$_x$CuO$_4$
Persisting Above the Bulk Superconducting Transition}

\author{J.E.~Sonier,$^1$ M.~Ilton,$^1$ V.~Pacradouni,$^1$ C.V.~Kaiser,$^1$
S.A. Sabok-Sayr,$^1$ Y.~Ando,$^2$ S.~Komiya,$^3$ W.N.~Hardy,$^4$ D.A.~Bonn$^4$, 
R.~Liang$^4$ and W.A.~Atkinson$^5$}

\affiliation{$^1$Department of Physics, Simon Fraser University, Burnaby, British Columbia V5A 1S6, Canada \\
$^2$Institute of Scientific and Industrial Research, Osaka University, Ibaraki, Osaka 567-0047, Japan \\
$^3$Central Research Institute of Electric Power Industry, Komae, Tokyo 201-8511, Japan\\
$^4$Department of Physics and Astronomy, University of British Columbia, Vancouver, British 
Columbia V6T 1Z1, Canada\\ 
$^5$Department of Physics and Astronomy, Trent University, Peterborough, Ontario K9J 7B8, Canada}


\begin{abstract}
We report that in YBa$_2$Cu$_3$O$_y$ and La$_{2-x}$Sr$_x$CuO$_4$ there is a 
spatially inhomogeneous response to magnetic field for temperatures $T$ extending well 
above the bulk superconducting transition temperature $T_c$.
An inhomogeneous magnetic response is observed above $T_c$ even in ortho-II YBa$_2$Cu$_3$O$_{6.50}$, 
which has highly ordered doping. The degree of the field inhomogeneity above $T_c$ tracks the 
hole doping dependences of both $T_c$ and the density of the superconducting carriers below $T_c$,
and therefore is apparently coupled to superconductivity.
\end{abstract}

\maketitle
The $T$-$p$-$H$ phase diagram of high-$T_c$ cuprates ({\it i.e.} temperature $T$, hole doping $p$, 
and magnetic field $H$) includes different magnetic phases discovered by the technique of 
muon spin rotation ($\mu$SR). As a sensitive local probe of static or quasistatic moments 
that are not necessarily ordered, $\mu$SR experiments have revealed that: 
(i) static Cu electronic moments remnant of the antiferromagnetic (AF) phase persist 
across the insulator-superconducting boundary \cite{Kiefl:89,Sanna:04}, 
(ii) some form of weak static magnetism occurs in YBa$_2$Cu$_3$O$_y$ (YBCO) 
near the pseudogap transition temperature $T^*$ \cite{Sonier:01}, and (iii) an 
applied magnetic field induces static magnetism in and around the vortex cores 
of samples on the low-doping side of 
what has been loosely dubbed a ``metal-to-insulator crossover'' (MIC) \cite{Sonier:07a}.           
Recently, (ii) has been independently verified by polar Kerr effect \cite{Xia:08} and
polarized neutron diffraction \cite{Fauque:06,Mook:08} measurements that
show the onset of some kind of magnetic order in YBCO near $T^*$. The neutron results 
indicate that the magnetic order is not AF Cu-spin order, but may be associated with
either the circulating-current phase proposed by Varma \cite{Varma}, a 
ferromagnetic arrangement of Cu spins, or the existence of staggered spins on the oxygen 
sites of the CuO$_2$ layers. From magnetic-field training of the polar Kerr effect
it is concluded in Ref.~\cite{Xia:08} that the magnetic order is 
associated with a time-reversal symmetry (TRS) breaking effect that 
persists above room temperature.       

We have used transverse-field muon-spin rotation (TF-$\mu$SR) to measure 
the local response in YBCO 
and La$_{2-x}$Sr$_x$CuO$_4$ (LSCO) single crystals to external magnetic fields up to $H \! = \! 7$~T. 
In this kind of 
experiment the initial muon spin polarization $P(t = 0)$ is oriented transverse to the field, 
which was applied perpendicular to the CuO$_2$ layers. The intrinsic spin of an implanted muon 
precesses in the plane perpendicular to the axis of the local magnetic field $B$ with a frequency 
$f_\mu \! = \! \gamma_\mu B$, where $\gamma_\mu \! = \! 851.6$~MHz/T is the muon's gyromagnetic ratio. 
Spatial field inhomogeneity in the bulk of the sample causes the transverse polarization $P(t)$ to decay 
with time due to the dephasing of muon spins precessing in different local magnetic fields.
Previously, we determined that none of the samples considered here contain Cu moments remnant of 
the AF phase that fluctuate slowly enough to be detected on the 
microsecond time scale of $\mu$SR \cite{Sonier:07a}. 
Furthermore, we showed that an applied magnetic field induces quasi-static magnetism 
for $T \! \ll \! T_c$, but only 
in samples below $y \! \approx \! 6.55$ for YBCO and $x \! \approx \! 0.16$ for LSCO. 

\begin{figure*}
\centering
\includegraphics[width=23.0cm]{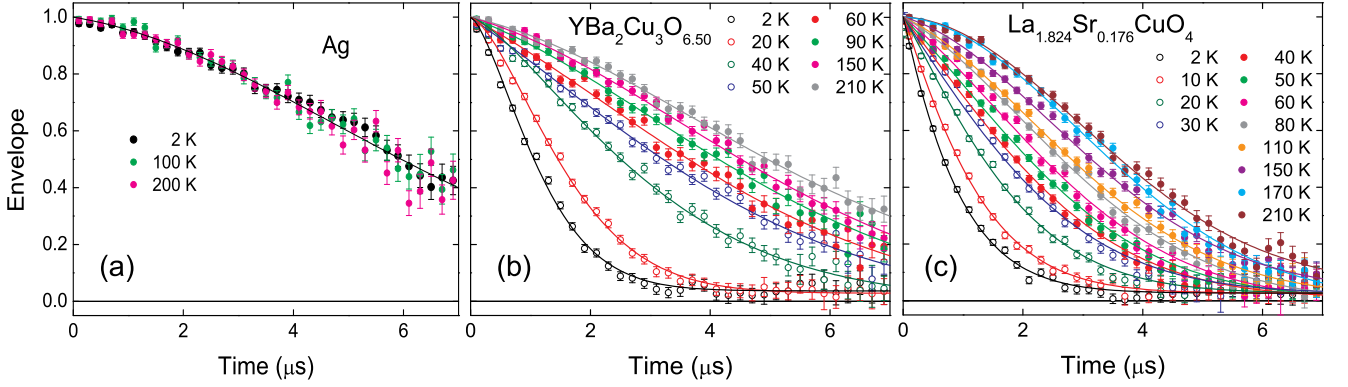}
\caption{(Color online) Representative envelopes of TF-$\mu$SR spectra at $H \! = \! 7$~T for  
(a) pure Ag, (b) YBCO with $y \! = \! 6.50$ ($T_c \! = \! 59$~K) and 
(c) LSCO with $x \! = \! 0.176$ ($T_c \! = \! 37.1$~K). The open and closed symbols in (b) and (c) 
indicate spectra for $T \! < \! T_c$ and $T \! > \! T_c$, respectively. The solid curves are the 
envelope function $G(t) \exp(-\Delta^2t^2)$ of Eq.~(\ref{eq:pol}).} 
\label{fig1}
\end{figure*}

The TF-$\mu$SR spectra are the sum of a sample signal and a time-independent background signal, 
$aP(t) \! = \! a_{\rm s} P_{\rm s}(t) \! + \! a_{\rm bg}$, where $a_{\rm bg}$ is less 
than 4~\% of the total signal amplitude $a$. The sample polarization 
function can be written as
\begin{equation}
P_{\rm s}(t) = G(t) \exp(-\Delta^2 t^2)\cos(f_\mu t + \phi) \, ,
\label{eq:pol}
\end{equation}
where the Gaussian function $\exp(-\Delta^2 t^2)$ accurately accounts for the random nuclear dipole fields 
and is temperature independent, and $G(t)$ is a phenomenological function that accounts for additional 
relaxation of $P_{\rm s}(t)$ by other sources. The relaxation function $G(t)\exp(-\Delta^2 t^2)$ is the 
``envelope'' of the TF-$\mu$SR signal.
Figure~1(a) shows TF-$\mu$SR envelopes for pure Ag, which does not exhibit superconductivity and does 
not contain electronic magnetic moments. The TF-$\mu$SR signals for Ag are well described by 
Eq.~(\ref{eq:pol}) with a {\it temperature-independent} exponential relaxation function 
$G(t) \! = \! \exp(-\Lambda_{\rm Ag}t)$, where $\Lambda_{\rm Ag}$ is a measure of the field inhomogeneity 
of the superconducting magnet used to generate the applied field. Alternatively, in YBCO and LSCO, the 
formation of a vortex lattice below $T_c$ creates a broad {\it temperature-dependent} internal magnetic 
field distribution $n(B)$. Below $T \! \approx \! 0.5 T_c$, the TF-$\mu$SR signals in YBCO and LSCO
are well described by Eq.~(\ref{eq:pol}) with a stretched-exponential relaxation function 
$G(t) \! = \! \exp[-(\Lambda t)^\beta]$, where $1.19 \! \leq \! \beta \! \leq \! 1.68$ for YBCO and 
$1.0 \! \leq  \! \beta \! \leq \! 1.19$ for LSCO. An exception is LSCO $x \! = \! 0.145$, which is 
discussed below. For $T \! > \! 0.5 T_c$, a single exponential relaxation function 
$G(t) \! = \! \exp(-\Lambda t)$ in Eq.~(\ref{eq:pol}) describes the TF-$\mu$SR signals for all samples.

\begin{figure}
\centering
\includegraphics[width=9.0cm]{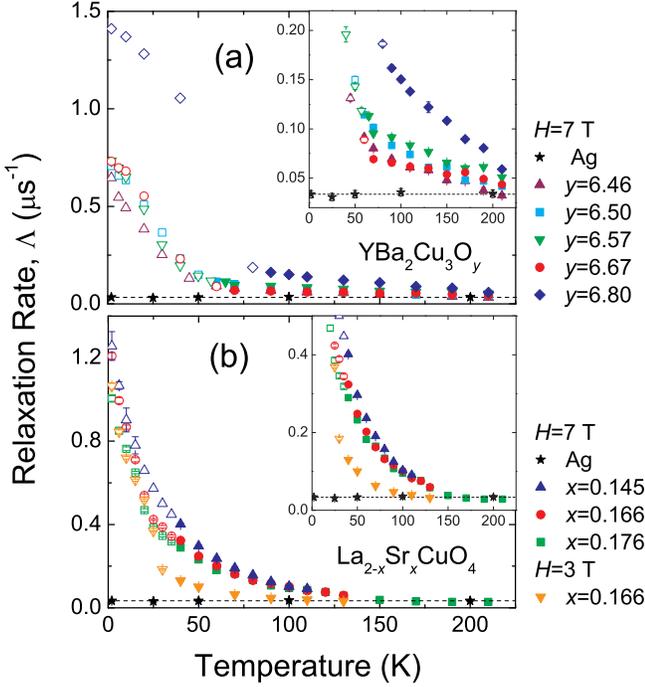}
\caption{(Color online) Temperature dependence of $\Lambda$ for (a) YBCO and (b) 
LSCO samples. The open and closed symbols correspond to data taken for $T \! < \! T_c$ and
$T \! > \! T_c$, respectively. Included in each panel is $\Lambda_{\rm Ag}$ for pure Ag. 
The insets are blowups of the main panels.}
\label{fig2}
\end{figure}

Above $T_c$ an applied magnetic field penetrates a nonmagnetic superconductor 
uniformly. Since $\Lambda$ is a measure of the width of $n(B)$ originating 
from sources other than nuclear dipoles, the difference $\Lambda - \Lambda_{\rm Ag}$ should vanish at 
$T_c$. Instead we find that a nonzero value of $\Lambda - \Lambda_{\rm Ag}$ persists to temperatures 
well above $T_c$ (see Fig.~\ref{fig2}). Note that above $T \! \approx \! 200$~K, the observed $\Lambda$ in 
YBCO is reduced by thermally activated hopping of the muon \cite{Sonier:02}. Hence the temperature at 
which $\Lambda - \Lambda_{\rm Ag}$ vanishes can only be determined by extrapolation.
We now discuss possible origins of the anomalous spatial field inhomogeneity above $T_c$. 

{\it Remnant Copper Spins}.---Previously 
Savici {\it et al.} \cite{Savici:05} detected field-induced inhomogeneity above $T_c$ 
in underdoped cuprates, including LSCO $x \! = \! 0.12$. In contrast to the current study, 
quasi-static Cu spins remnant of the AF phase were detected for $H \! = \! 0$. Field-induced 
ordering of these spins in a significant volume of the sample adds a second faster-relaxing 
exponential component to $G(t)$. This was observed in LSCO $x \! = \! 0.12$ \cite{Savici:05}, 
but also here in LSCO $x \! = \! 0.145$---{\it i.e.} on the low-doping side of the MIC near 
$x \! = \! 0.16$. However, in both cases, this second component vanishes well below $T_c$, 
as does disordered static magnetism induced at lower fields \cite{Sonier:07a}. We note that
neutron scattering measurements on LSCO $x \! = \! 0.10$ also show field-induced magnetic
order only below  $T_c$ \cite{Lake:02}, and measurements on LSCO $x \! = \! 0.163$
show no static magnetism up to $H \! = \! 14.5$~T and dynamic AF correlations 
only below $T_c$ \cite{Lake:01}.

\begin{figure}
\centering
\includegraphics[width=14.0cm]{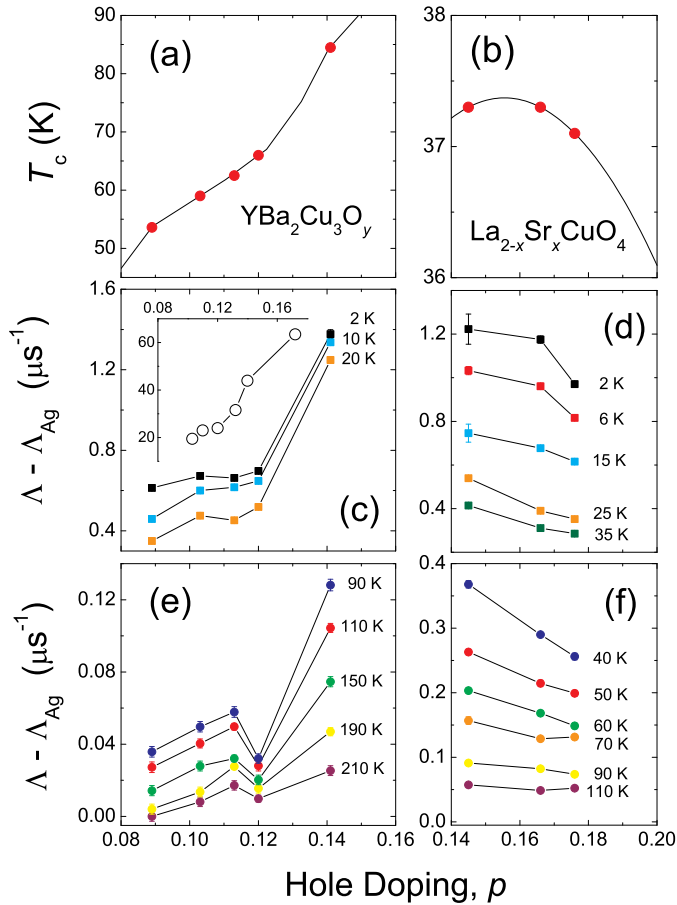}
\caption{(Color online) Hole doping dependence of $T_c$ in (a) YBCO and (b) LSCO for $H \! = \! 7$~T. 
The values of $p$ for LSCO correspond to the Sr concentration, whereas $p$ for YBCO was determined 
in Ref.~\cite{Liang:06}. The relaxation rate $\Lambda \! - \! \Lambda_{\rm Ag}$ for 
YBCO is plotted versus $p$ for (c) $T \! < \! T_c$ and (e) $T \! > \! T_c$, and for LSCO 
for (d) $T \! < \! T_c$ and (f) $T \! > \! T_c$. 
The inset of (c) shows the hole doping dependence of $1/\lambda_{ab}^2$ (in $\mu$m$^{-2}$) in YBCO 
\cite{Sonier:07b}.}
\label{fig3}
\end{figure}

Figure~\ref{fig3} shows the hole-doping dependence of $\Lambda \! - \! \Lambda_{\rm Ag}$. 
Despite YBCO $y \! = \! 6.67$ being close to 1/8 hole doping where spin/charge stripe 
correlations \cite{Kivelson:03} and a suppression of superconductivity \cite{Sonier:07b} occur, 
above $T_c$ the value of $\Lambda \! - \! \Lambda_{\rm Ag}$ is smaller 
than that for the $y \! = \! 6.50$, 6.57 and 6.80 samples. Furthermore, the largest field inhomogeneity 
detected in YBCO above $T_c$ is in the $y \! = \! 6.80$ sample, which is furthest away from the AF phase. 
In other words, the hole doping dependence of $\Lambda - \Lambda_{\rm Ag}$ presented
in Fig.~\ref{fig3}(e) is inverted from the usual tendency for the field inhomogeneity 
due to remnant static or fluctuating Cu spins.

For $T \! < \! T_c$, $\Lambda \! - \! \Lambda_{\rm Ag}$ tracks the inhomogeneous 
field of the vortex lattice and 
hence tracks the superfluid density $n_s \! \propto \! 1/\lambda_{ab}^2$, 
where $\lambda_{ab}$ is the in-plane magnetic penetration depth 
[see inset of Fig.~\ref{fig3}(c)].  Since $T_c$ varies as a function of $n_s$ \cite{Uemura:89}, 
we see in Figs.~\ref{fig3}(c) and \ref{fig3}(d) that $\Lambda \! - \! \Lambda_{\rm Ag}$ also tracks $T_c$. 
Remarkably this continues to be the case well above $T_c$ [Figs.~\ref{fig3}(e) and \ref{fig3}(f)].  
It is thus evident that nonzero $\Lambda \! - \! \Lambda_{\rm Ag}$ above $T_c$ is in some way 
coupled to superconductivity.

\begin{figure}
\centering
\includegraphics[width=9.0cm]{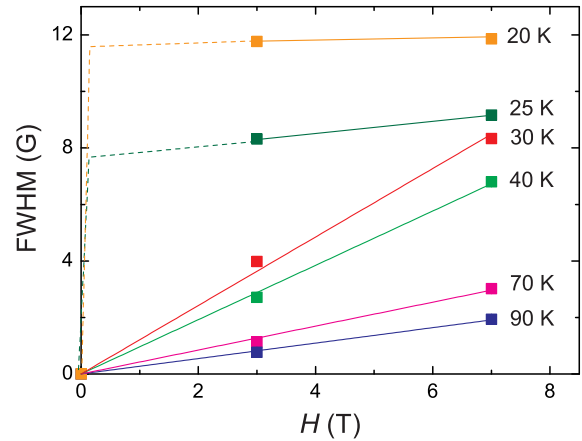}
\caption{(Color online) Field dependence of the full width at half maximum 
(FWHM = $2(\Lambda \! - \! \Lambda_{\rm Ag})/\gamma_\mu$) of the Lorentzian magnetic field distribution 
corresponding to the exponential relaxation for LSCO $x \! = \! 0.166$ 
($T_c \! = \! 37.3$~K).}
\label{fig4}
\end{figure}

{\it Vortex Liquid}.---Ong {\it et al.} \cite{Wang:06} have established that 
the application of a sizeable field creates 
a Nernst effect and a corresponding field-enhanced diamagnetic signal indicative of a two-dimensional
(2D) vortex liquid, which persists 
above $T_c$, but is contained within the more extensive pseudogap region.
In cuprates, vortices fluctuate about their equilibrium positions with a 
characteristic fluctuation time ($\sim \! 10^{-11}$~s \cite{Brandt:89}) that is much smaller than 
the timescale ($10^{-6}$~s) of $\mu$SR, with a fluctuation-amplitude in the liquid phase on the 
order of the intervortex spacing. These factors conspire to produce severe motional narrowing of the 
field distribution detected by $\mu$SR \cite{Brandt:91}, as observed in 
Bi$_2$Sr$_2$CaCu$_2$O$_{8+\delta}$ (BSCCO) at $T \! < \! T_c$ and $H \! \leq \! 0.1$~T \cite{Lee:95}. 
While YBCO and LSCO are less anisotropic than BSCCO, vortex-lattice melting and a loss of 
vortex line tension also occur before $T_c$ is reached \cite{Figueras:06}. As the external magnetic 
field is increased, a signature of a 2D vortex liquid is a reduction of the 
$\mu$SR line width \cite{Brandt:91,Lee:95}. However, here we observe the exact opposite. As shown in
Figs.~\ref{fig2}(b) and \ref{fig4}, $\Lambda$ {\it increases} with increasing field for $T \! > \! T_c$ 
and remains nonzero even beyond $T \! = \! 2T_c$. 
Thus the field inhomogeneity detected above $T_c$ cannot be caused by vortices, which is not to 
say a vortex liquid does not exist. 

\begin{figure}
\centering
\includegraphics[width=12.0cm]{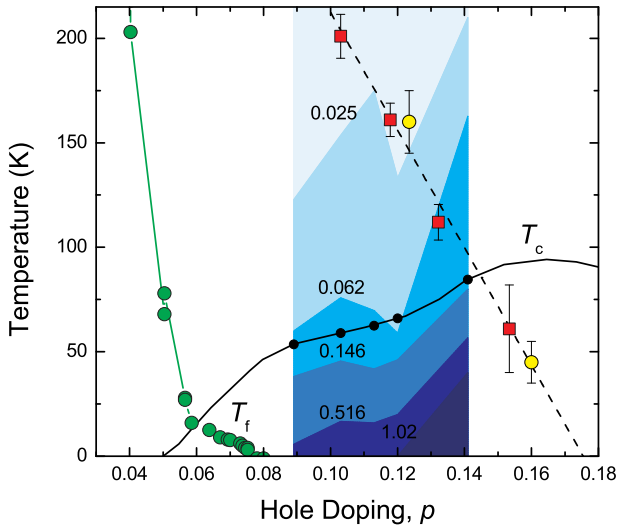}
\caption{(Color online) Contours of the magnitude of $\Lambda \! - \! \Lambda_{\rm Ag}$
for YBCO at $H \! = \! 7$~T. Also shown is the extended spin freezing transition $T_f$ determined by 
ZF-$\mu$SR \cite{Sanna:04}, and the onset of magnetic order near $T^*$ detected 
by ZF-$\mu$SR \cite{Sonier:01} (yellow circles) and polar Kerr effect \cite{Xia:08} (red squares)
measurements. The ZF-$\mu$SR data of Ref.~\cite{Sonier:01} has been corrected using the
appropriate $T_c$ to $p$ conversion for YBCO \cite{Liang:06}.}
\label{fig5}
\end{figure}

{\it Inhomogeneous Superconductivity}.---It has been proposed that
superconductivity first develops in small domains at $T^*$ \cite{Alvarez:01,Kresin:06,Mello:07}
that proliferate with decreasing temperature, eventually resulting in the formation 
of the bulk-superconducting phase at $T_c$ via percolation or Josephson coupling.   
This picture is supported by the recent detection of spatially inhomogeneous 
pairing gaps in BSCCO above $T_c$ \cite{Gomes:07}, and is compatible with   
the hysteresis observed in low-field magnetization measurements on LSCO
above $T_c$ by Panagopoulos {\it et al.} \cite{Pano:06}. 

Geshkenbein {\it et al.} \cite{Geshkenbein:98} have shown that the presence
of small superconducting regions with $T_c$ greatly exceeding the bulk $T_c$
explains an unusual linear diamagnetic response ($M \! \sim \! H$) observed
above the resistive transition of highly-overdoped Tl$_2$Ba$_2$CuO$_{6+\delta}$  
\cite{Bergemann:98}. In their model, the magnetization of a superconducting grain
is proportional to $H$. Because $\Lambda/H$ is proportional to the spread in local
magnetic susceptibility at the muon site, then a signature of inhomogeneous 
superconductivity is a linear dependence of $\Lambda$ on $H$.
As shown in Fig.~\ref{fig4} this is observed
for $T \! > \! 0.8$~$T_c$, and is distinct from the behavior 
at lower $T$ where the width of $n(B)$ is dominated by the field inhomogeneity 
of the vortex lattice. This was also observed in Ref.~\cite{Savici:05}. 
The onset temperature for 
$\Lambda \! \propto \! H$ behavior is compatible with a 
thermally induced break-up of the bulk superconducting state into 
small superconducting domains, beginning at a temperature slightly below $T_c$.
Nevertheless, this interpretation must be regarded as speculative,
since we cannot say whether there is a diamagnetic response. Also,
while electronic phase separation in the form of domains is understandable
in cuprates doped by cation substitution, such as BSCCO 
and LSCO, it is not clear why this should be equally prevalent in YBCO where 
doping occurs via a change in the oxygen concentration of the CuO-chain layers. 
Of particular note is ortho-II YBCO $y \! = \! 6.50$, which has alternating full and 
empty CuO chains, and therefore has highly ordered doping. Even so, 
the degree of field inhomogeneity in YBCO is comparable to LSCO, as indicated 
by the similar values of $\Lambda \! - \! \Lambda_{\rm Ag}$ at $p \! \approx \! 0.14$ and 
$T \! \geq \! 90$~K [Figs.~3(e) and 3(f)].

We conclude by showing that the spatially inhomogeneous response of YBCO to a $7$~T 
field persists above the anomalous magnetic order that occurs in zero field
near $T^*$ (Fig.~\ref{fig5}). If the response is due to electronic moments, 
what is novel here is that it tracks superconductivity. A compatible picture is one 
in which {\it magnetic} normal-state carriers form pairs and condense at $T_c$.
Alternatively, this behavior could arise from spin magnetism being confined to
isolated regions by inhomogeneous superconductivity.   
   
We thank S. A. Kivelson, E. H. Brandt, G. I. Menon, A. J. Millis, G. M. Luke and 
Y. J. Uemura for helpful discussions, R. Kadono for sharing beam time, and 
TRIUMF for technical assistance. 
This work was supported by the Natural Sciences and Engineering Research Council of 
Canada, the Canadian Institute for Advanced Research, and in Japan by KAKENHI 19674002.

\end{document}